\documentclass[conference]{IEEEtran} 
\IEEEoverridecommandlockouts

\usepackage{amssymb}
\usepackage{amsmath}
\usepackage{amsmath,bm}
\usepackage{amsthm}
\usepackage{graphicx}
\usepackage{subfigure}
\usepackage{cite}
\usepackage{enumerate}
\usepackage{algorithm}
\usepackage{algorithmicx}
\usepackage{algpseudocode}
\usepackage{color, soul}
\usepackage{booktabs}
\usepackage{epstopdf}
\let\labelindent\relax
\usepackage{enumitem}
\usepackage{color}

\begin{document}

\newtheorem{definition}{\bf Definition}
\newtheorem{observation}{\bf Observation}
\newtheorem{proposition}{\bf Proposition}
\newtheorem{theorem}{\bf Theorem}
\newtheorem{remark}{\bf Remark}

\renewcommand{\algorithmicrequire}{\textbf{Input:}} 
\renewcommand{\algorithmicensure}{\textbf{Output:}} 

\title{\Large{Sense-Store-Send: Trajectory Optimization for a Buffer-aided Internet of UAVs}}

\author{
{Yujie Jin},
{Hongliang Zhang}, \IEEEmembership{Member, IEEE},
{Shuhang Zhang}, \IEEEmembership{Student Member, IEEE},\\
{Zhu Han}, \IEEEmembership{Fellow, IEEE},
{and Lingyang Song}, \IEEEmembership{Fellow, IEEE}

\thanks{Y. Jin, S. Zhang, and L. Song are with Department of Electronics, Peking University, Beijing, China  (Email: \{jyj17pku, shuhangzhang,lingyang.song\}@pku.edu.cn).}

\thanks{H. Zhang is with Department of Electronics, Peking University, Beijing, China, and also with Electrical and Computer Engineering Department, University of Houston, Houston, TX, USA (Email: hongliang.zhang92@gmail.com).}


\thanks{Z. Han is with Electrical and Computer Engineering Department, University of Houston, Houston, TX, USA, and also with the Department of Computer Science and Engineering, Kyung Hee University, Seoul, South Korea (Email: zhuhan22@gmail.com).}
\vspace{-6mm}}
\maketitle

\begin{abstract}
In this letter, we study a buffer-aided Internet of unmanned aerial vehicles (UAVs) in which a UAV performs data sensing, stores the data, and sends it to the base station~(BS) in cellular networks. To minimize the overall completion time for all the sensing tasks, we formulate a joint trajectory, sensing location and sensing time optimization problem. To solve this NP-hard problem efficiently, we propose an iterative trajectory, sensing location and sensing time optimization (ITLTO) algorithm, and discuss the trade-off between sensing time and flying time. Simulation results show that the proposed algorithm can effectively reduce the completion time for the sensing tasks.
\end{abstract}
\vspace{-4mm}
\section{Introduction}
\vspace{-2mm}
In recent years, as an emerging technology, unmanned aerial vehicles (UAVs) are widely applied in military, public and civil applications. Among these applications, the use of UAVs to perform sensing tasks attracts the interest of industry and academia due to the advantages of high mobility, low cost, and large service coverage~\cite{ZSH2019}. In some real-time applications, such as traffic monitoring, sensory data needs to be sent to the base station (BS) through cellular networks for further processing, which are referred to as Internet of UAVs.

Some of the initial works have studied the UAV sensing and sending optimization in the Internet of UAVs. In~\cite{WQQ2018}, the authors minimized the total flight time of a UAV which moves along a straight line to collect data from a set of sensors. In~\cite{YJD2019}, the authors minimized the completion time for all the tasks in a cooperative Internet of UAVs. Most of existing works assume that UAVs are capable to perform real-time data transmission. However, in case of poor wireless signal quality or no need for real-time sending, the data of multiple tasks can be stored and uploaded uniformly, which can reduce the time to complete all the tasks. This can be achieved by deploying the buffer on the UAV.


In this letter, we study a buffer-aided Internet of UAVs in which one UAV senses data from a series of  tasks, stores data in a buffer, and sends data to a BS. We consider the generation of sensory data as a random process and establish a probabilistic model for the UAV's joint sensing, storing and sending  process. To minimize the completion time of the UAV for all the sensing tasks while guaranteeing successful sensing probability, we formulate the UAV trajectory optimization problem. We propose an iterative trajectory, sensing location and sensing time optimization (ITLTO) algorithm to solve the problem efficiently. Unlike existing works that require the UAV to complete the sending of a task before the sensing of the next one, a buffer enables the UAV to transmit the collected data of previous tasks while sensing the current target, which brings more freedom on the scheduling of sensing and sending. However, the deployment of buffer may bring some delay of data packets since the collected data is not transmitted immediately, and thus the buffer-aided Internet of UAVs is more suitable for delay-tolerant applications.

\vspace{-2mm}

\begin{figure}[!t]
\centerline{\includegraphics[width=3.0in]{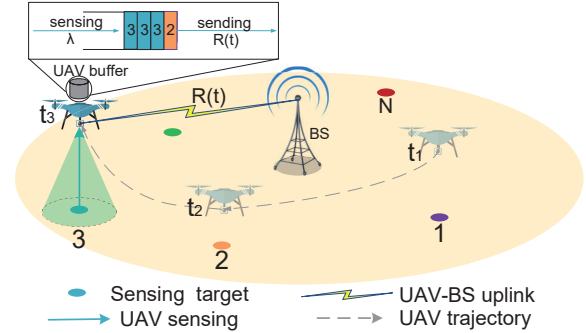}}\vspace{-3mm}
\caption{System model of a buffer-aided Internet of UAVs.}
\vspace{-5mm}
\label{systemmodel}
\vspace{-1mm}
\end{figure}
\section{System Model}
\vspace{-2mm}
\label{System}
We consider a buffer-aided Internet of UAVs\footnote{In this letter, we assume that the Internet of UAVs uses orthogonal frequency multiple access (OFDMA) to support multiple UAVs. Since the trajectory optimization for each UAV is independent, we only consider one UAV for brevity.} as shown in Fig.~\ref{systemmodel}, which consists of one BS and one UAV to execute $N$ sensing tasks within the cell coverage, denoted by $\mathcal N$ = $\left\{ 1,2,...,N  \right\}$. The UAV is required to collect the sensory data for $N$ sensing tasks in a given order\footnote{In this letter, we assume that the ordering of the tasks is known before the trajectory optimization, which can be obtained by the algorithm proposed in~\cite{ZZZ2018}.}, and upload the data to the BS for further processing. The sensory data can be stored in a capacity-limited UAV buffer until it is sent to the BS. In the following, we introduce UAV sensing, UAV storing, and UAV sending in detail.

\subsection{UAV Sensing}
\vspace{-1mm}
We assume that each task contains only one target to be sensed by the UAV. In the following, we refer to the target of the $n$-th sensing task as target $n$. In the UAV sensing procedure, the UAV first moves to a sensing point, and then hovers for a number of time slots to collect the data.\footnote{We call the location that the UAV hovers to sense target $n$ as UAV sensing location $n$. We assume the UAV reaches sensing location $n$ in time slot $t_n$.} Without loss of generality, we denote the location of the BS by $(0,0,H)$, and the location of target $n$ by $(x_n,y_n,0)$. In time slot $t$, let $\bm{l}(t)=(x(t),y(t),z(t))$ be the location of the UAV, and $\bm{v}(t)=(v_x(t),v_y(t),v_z(t))$ be its velocity, with $\bm {v}(t)=\bm{l}(t)-\bm{l}(t-1)$. Due to the mechanical limitation, the speed of the UAV is no more than $v_{max}$.

We utilize the probabilistic sensing model to quantify the performance of the sensing procedure~\cite{ZSH2019}, where the successful sensing probability is an exponential function of the distance between the UAV and the target. To be specific, the successful sensing probability for task $n$ can be expressed as
\begin{equation}\label{successful sensing probability}\vspace{-2mm}
PR_n=e^{-\nu d_n},
\end{equation}
where $\nu$ is a parameter evaluating the sensing performance, and $d_n$ is the distance between the UAV sensing location $n$ and target $n$, which can be given by
\begin{equation}\label{distance UT}\vspace{-2mm}
d_n=\sqrt{(x(t_n)-x_n)^2+(y(t_n)-y_n)^2+(z(t_n))^2}.
\end{equation}

\subsection{UAV Storing}
\vspace{-1mm}
When the UAV performs sensing for task $n$, we assume that the generation of sensory data obeys the Poisson process with the rate of $\lambda$~\cite{LH2017}. The size of each data packet has a general distribution with an average size of $R_s$. The collected data is stored in the UAV buffer until it is sent to the BS. Let $Q(t)$ be the amount of collected data in the buffer in time slot $t$. Since the capacity of the buffer is limited, the number of data packets in the buffer cannot be more than $K$, i.e.,
\begin{equation}\label{buffer capacity}\vspace{-2mm}
0\leq Q(t)\leq K\times R_s.
\end{equation}

\subsection{UAV Sending}
\vspace{-1mm}
For simplicity, it is assumed that each UAV is assigned with a dedicated subchannel in the system, and thus there is no interference in UAV sending.\footnote{The UAV can be supported by orthogonal frequency multiple access (OFDMA) or nonorthogonal multiple access (NOMA)~\cite{JWY2020}.}  We utilize the air-to-ground transmission channel model proposed in~\cite{ZSH2019}. The average pathloss in dB can be expressed as
\begin{equation}\label{average PL}\vspace{-2mm}
PL_{a}(t)=P_{L}(t)\times PL_{L}(t)+P_{N}(t)\times PL_{N}(t),
\end{equation}
where $PL_{L}(t)$ is the line-of-sight (LoS) pathloss, $PL_{N}(t)$ is the none-line-of-sight (NLoS) pathloss, $P_{L}(t)$ is the probability of LoS connection, and $P_{N}(t)$ is that of NLoS connection.

As the antennas on the UAV and the BS are placed vertically, the probability of LoS connection is given by $P_{L}(t)=\left({1+\alpha\exp(-\beta(\phi(t)-\alpha))}\right)^{-1},$ where $\alpha$ and $\beta$ are environmental parameters, and $\phi(t)=\sin^{-1}((z(t)-H)/d_{U,B}(t))$ is the elevation angle. The average received power of the BS from the UAV is given by
\begin{equation}\label{BS receive power}\vspace{-2mm}
P_{R}(t)={P_T(t)}/{10^{PL_{a}(t)/10}},
\end{equation}
where $P_T(t)$ is the transmission power of the UAV in time slot $t$, which is considered as a fixed value in this letter. The data rate from the UAV to the BS in time slot $t$ is
\begin{equation}\label{rate}\vspace{-2mm}
R(t)=W_B\times\log_2\left(1+P_{R}(t)/\sigma^2\right),
\end{equation}
where $W_B$ is the bandwidth of the subchannel, and $\sigma^2$ is the variance of additive white Gaussian noise with zero mean. The change of data in the buffer during the joint sensing, storing, and sending process will be discussed in the following section. When the UAV is not performing sensing, the remain data in the buffer over time can be simply given by
\begin{equation}\label{change of data}\vspace{-1mm}
Q(t+1)=\max\left\{Q(t)-R(t),0\right\}.
\end{equation}

\section{Problem Formulation}
\vspace{-1mm}
\label{Problem}
In this section, we first introduce the concept of the task completion probability to evaluate the performance of executing a task. Afterwards, based on the task completion probability, we introduce the concept of task completion time, and then formulate a task completion time minimization problem.
\subsection{Task Completion Probability}
\vspace{-1mm}
When the UAV performs sensing and sending concurrently for task $n$, we assume that the data generation obeys Poisson process and the sending time for each packet is a constant $X_n=\frac{R_s}{R(t_n)}$ in time slot $t_n$. Thus, an $M/G/1/K$ queuing system~\cite{GH1985} which contains a single server and a finite-capacity queue with Poisson arrivals, can be used to model the joint sensing, storing and sending  in the proposed system.

Let $\pi_i$ be the probability of $i$ packets in the buffer after the steady state is reached. According to the results in~\cite{JM2011}, we have
\vspace{-1mm}
\begin{equation}\label{state probability}
\begin{aligned}
&\pi_0=\frac{\rho_n-1}{\rho_n^{2(K+1-\sqrt{\rho_n})/(2-\sqrt{\rho_n})}-1},\\
&\pi_K=\frac{\rho_n^{(2K-\sqrt{\rho_n})/(2-\sqrt{\rho_n})}(\rho_n-1)}{\rho_n^{2(K+1-\sqrt{\rho_n})/(2-\sqrt{\rho_n})}-1},\\
&\pi_i=\frac{\rho_n^{i-1}(\rho_n-1)}{\rho_n^{K-1}-1}(1-\pi_0-\pi_K),~i=1,\ldots,K-1,
\end{aligned}
\end{equation}
where $\rho_n=\lambda X_n$ is the traffic load offered to the queue. For each task, we assume that it is completed when the UAV successfully senses at least $C$ data packets. Let $C_n$ be the minimum number of packets that the UAV needs to collect for task $n$, which is inversely proportional to its successful sensing probability,\footnote{Since the UAV cannot determine whether the sensed data is successful or not due to its limited computational capability, we use the expectation to define the minimum number of packets.} i.e., $C_n = C/PR_n$.

Let $\delta_n$ be the UAV sensing time, i.e., the number of time slots that the UAV spends on sensing  target $n$.  We call task $n$ is \emph{completed} in time slot $t_n+\delta_n$ if the number of packets stored into the buffer during $\delta_n$ time slots reaches a threshold $C_n$. The change of data in the buffer during $\delta_n$ time slots is caused by two factors: the  generated data of the sensing task, and the data sent to the BS, i.e., $R(t_n)\times \delta_n$. Thus, task $n$ is completed in time slot $t_n+\delta_n$ if there are no less than $\lceil C_n+\frac{Q(t_n)-R(t_n)\times\delta_n}{R_s}\rceil$ packets existing in the buffer in this time slot. Accordingly, the task completion probability for task $n$ in time slot $t_n+\delta_n$ is expressed by
\begin{equation}\label{task completion probability}\vspace{-1mm}
p_n=\sum_{i=\lceil C_n+\frac{Q(t_n)-R(t_n)\times\delta_n}{R_s}\rceil}^{i=K}\pi_i.
\end{equation}

To guarantee the sensing performance of each task, the task completion probability for each task should be no less than a threshold $p_{min}$, i.e., $p_n\geq p_{min}$.
\vspace{-1mm}

\subsection{Completion Time Minimization}
\vspace{-1mm}
In the system, the task completion time is defined as the number of time slots that the UAV spends on completing the sensing and sending for all the tasks, i.e.,
\vspace{-1mm}
\begin{equation}\label{completion time}
T=t_N+\delta_N+T_{res},
\end{equation}
where $T_{res}$ is the time that the UAV spends on sending the residual data in the buffer after completing sensing for the last task $N$. We aim to minimize the task completion time $T$ by jointly optimizing the UAV trajectory $\bm{v}(t)$, the UAV sensing locations $\bm{l}(t_n)$, and UAV sensing time $\delta_n$ for all the tasks. Thus, the problem can be written by
\vspace{-2mm}
\begin{subequations} \label{system_optimization}
\begin{align}
&\mathop{\min}\limits_{\textbf{$\{\bm{v}(t)\}\{\bm{l}(t_n)\}\{\delta_n\}$}}T \label{Obj}\\
\textbf{\emph{s.t. }}
&0\leq Q(t)\leq K \times R_s,\label{system_1}\\
&p_n\geq p_{min},\label{system_2}\\
&\|\bm{v}(t)\|\leq v_{max},\label{system_3}\\
&\|\bm{v}(t)\|=0, t_n\leq t\leq t_n+\delta_n.\label{system_4}
\end{align}
\end{subequations}
Constraints on the buffer capacity, task completion probability and UAV speed are given in~(\ref{system_1}),~(\ref{system_2}) and~(\ref{system_3}), respectively. (\ref{system_4}) shows that the UAV velocity is zero when performing data sensing.

\subsection{Problem Decomposition}
\vspace{-1mm}
Problem~(\ref{system_optimization}) contains continuous variables $\bm{v}(t)$ and $\bm{l}(t_n)$, and integer variable $\delta_n$, which is NP-hard. To solve this problem efficiently, we propose an ITLTO algorithm, which decomposes the problem into two subproblems: 1) UAV trajectory optimization; and 2) UAV sensing location and sensing time optimization, and solve them iteratively.

In the UAV trajectory optimization subproblem, given the UAV sensing locations $\bm{l}(t_n), \forall n \in \mathcal N$ and the UAV sensing time $\delta_n, \forall n \in \mathcal N$, we aim to minimize the UAV flying time, which is defined as the time that the UAV spends on flying between two successive sensing locations. Without loss of generality, we study the trajectory between tasks $n$ and $n+1$. The UAV trajectory optimization subproblem can be written as:
\begin{equation} 
\begin{aligned}
&\mathop{\min}\limits_{\textbf{$\{\bm{v}(t)\}$}}t_{n+1}-(t_n+\delta_n) \\
\textbf{\emph{s.t. }}
&\sum_{t=t_n+\delta_n}^{t=t_{n+1}}\bm{v}(t)=\bm{l}(t_{n+1})-\bm{l}(t_n),\\
&\text{(\ref{system_1}), (\ref{system_2}), and (\ref{system_3})}.
\label{trajectory_optimization}
\end{aligned}
\end{equation}

In the UAV sensing location and sensing time optimization subproblem, given the UAV trajectory optimization result, we aim to minimize the completion time by adjusting the sensing locations and UAV sensing time. The subproblem can be written as:
\begin{equation} \label{UAV sensing location and sensing time optimization formulation}
\begin{aligned}
&\mathop{\min}\limits_{\textbf{$\{\bm{l}(t_n)\}\{\delta_n\}$}}T \\
\textbf{\emph{s.t. }}
&\text{(\ref{system_1}) and (\ref{system_2})}.
\end{aligned}
\end{equation}

\section{Iterative Trajectory, Location, and Time Optimization Algorithm}
\vspace{-1mm}
\label{Algorithm}
In this section, we first propose an iterative algorithm that solves the two subproblems, respectively.\footnote{It is worthwhile to mention that the proposed algorithm is an offline one, and thus, we only consider the pathloss in this letter.} Then we discuss the convergency and complexity of the algorithm, and the trade-off between UAV flying time and UAV sensing time.
\vspace{-1mm}
\subsection{UAV Trajectory Optimization}\label{Trajectory}
\vspace{-1mm}
Note that if the UAV speed is given, the line segment between two successive tasks corresponds to the minimum UAV flying time. To minimize the completion time, we set $\bm{l}(t_{n+1})-\bm{l}(t_n)$ as the UAV's moving direction between tasks $n$ and $n+1$. In the following, we will derive the optimal speed.

Since time is discrete and problem~(\ref{trajectory_optimization}) is hard to be solved directly due to the complicated expressions, we solve the problem by enumerating the UAV flying time $T_n^{f}=t_{n+1}-(t_n+\delta_n)$. In each enumeration, we first remove constraint~(\ref{system_2}) and maximize the sending data rate in each time slot. Thus, the problem can be written as
\begin{subequations} \label{enumeration problem}
\begin{align}
&\mathop{\max}\limits_{\textbf{$\{{v}(t)\}$}}R(t) \label{system 15}\\
\textbf{\emph{s.t. }}
&0\leq \|\bm{v}(t)\|\leq v_{max},\label{system 18}\\
&\sum_{t=1}^{t=T_n^{f}}{v}(t)=L=\parallel \bm{l}(t_{n+1})-\bm{l}(t_n)\parallel.\label{system 19}
\end{align}
\end{subequations}

We denote the length that the UAV has moved before time slot $t$ by $L(t)$. To satisfy constraints~(\ref{system 18}) and~(\ref{system 19}), the feasible range of the UAV speed in time slot $t$ is $max\left\{0,L-L(t)-v_{max}\times (T_n^{f}-t-1)\right\}\leq v(t)\leq v_{max}$. Problem~(\ref{enumeration problem}) can be converted to:

\vspace{-4mm}
\begin{small}
\begin{subequations} \label{converted enumeration problem}
\begin{align}
&\mathop{\max}\limits_{\textbf{$\{{v}(t)\}$}}R(t) \label{system 20}\\
\textbf{\emph{s.t. }}
&\max\left\{0,L-L(t)-v_{max}\times (T_n^{f}-t-1)\right\}\leq v(t)\leq v_{max}.\label{system 21}
\end{align}
\end{subequations}
\end{small}
\vspace{-4mm}

According to the results in~\cite{ZZD2019}, the pathloss variables $PL_L(t)$ and $PL_N(t)$ change much faster than the LoS probability variables $P_L(t)$ and $P_N(t)$ with the movement of the UAV. Thus, the LoS probability can be regarded as a constant in a single time slot and the sending rate is only determined by pathloss $PL_L(t)$ and $PL_N(t)$. Therefore, problem~(\ref{converted enumeration problem}) is approximated as a convex problem and can be solved by existing optimization methods.

In each enumeration, we solve problem~(\ref{converted enumeration problem}) and check whether the optimal solution satisfies constraint~(\ref{system_2}). If the solution is found, the enumeration terminates and the current $T_n^{f}$ is the optimal solution to problem~(\ref{trajectory_optimization}). Otherwise, we let $T_n^{f}=T_n^{f}+1$ and repeat the above processes.
\vspace{-1mm}

\subsection{UAV Sensing Location And Sensing Time Optimization}
\vspace{-1mm}
In this section, we introduce a method to optimize the UAV sensing locations and sensing time. If the UAV trajectory and sensing locations are given, the only constraint of the UAV sensing time is determined by the inequality~(\ref{system_2}) and the lower bound of $\delta_n$ can be solved by
\vspace{-2mm}
\begin{equation}\label{sensing time optimization}
\delta_n=\min\left\{\left\lceil\frac{R_s}{R(t_n)}\left[C_n+\frac{Q(t_n)}{R_s}-1-\left\lfloor\frac{\ln\rho^{*}}{\ln\rho_n}\right\rfloor\right]\right\rceil, 1\right\},
\end{equation}
where $\rho^{*}=\frac{(1-\rho_n^{K-1})(p_{min}-\pi_K)}{1-\pi_0-\pi_K}+\rho_n^{K-1}$. We can optimize the UAV sensing time as~(\ref{sensing time optimization}) after determining the UAV sensing locations. When optimizing sensing location $n+1$, we fix the UAV flying time between tasks $n$ and $n+1$, and the corresponding UAV speed as the results obtained in Section \ref{Trajectory}, which are denoted by $T_n^{f}$ and $\left\{\bm{v}(1), \bm{v}(2),...,\bm{v}(T_n^{f}) \right\}$, respectively. Then, We use a local search method to solve this problem as elaborated below:

\textbf{Step 1:}  We adjust $\bm{l}(t_{n+1})=\bm{l}(t_{n})+\sum_{i=1}^{i=T_n^{f}}\bm{v}(i)$ through increasing $T_n^{f}$ by one time slot with the UAV speed being $\bm{v}_{max}$, or reducing $T_n^{f}$ by one time slot with the UAV speed being $\bm{v}(T_n^{f})$.

\textbf{Step 2:} We check whether the local adjustment can reduce the task completion time of task $n+1$. If so, we update the solution, and otherwise, we keep the previous solution.
\vspace{-1mm}

\subsection{Overall Algorithm}
\begin{figure}[!t]
\centerline{\includegraphics[width=3.4in]{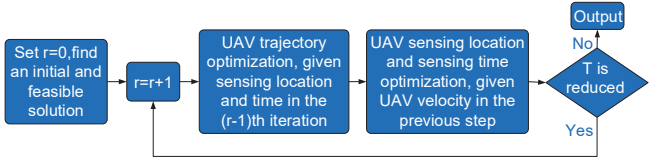}}\vspace{-3mm}
\caption{Flowchart of ITLTO algorithm.}
\vspace{-5mm}
\label{flowchart}
\vspace{-2mm}
\end{figure}

\vspace{-1mm}
The ITLTO algorithm is summarized as follows: We first find a feasible solution to problem~(\ref{system_optimization}) that satisfies its all constraints. Then, we perform the two optimization subproblems iteratively, until the task completion time is converged. In each iteration, we first perform UAV trajectory optimization given the UAV sensing locations and sensing time obtained in the last iteration. Then, we perform UAV sensing location and sensing time optimization, given the UAV flying time and speed among tasks obtained in the UAV trajectory optimization. We give a flowchart to summarize the ITLTO algorithm in Fig.~\ref{flowchart}.
\vspace{-2mm}

\subsection{Performance Analysis}\label{analysis}
\vspace{-1mm}
In the following, we first provide two propositions on the convergence and complexity. Then, we discuss the relation between the sensing time and the total completion time.

\textbf{Proposition 1:} The ITLTO algorithm is convergent.

\begin{proof}
In the UAV trajectory optimization, given the UAV sensing time in the last iteration, we minimize the UAV flying time and the task completion time decreases with the UAV trajectory optimization. In the UAV sensing location and sensing time optimization, the completion time of each task does not increase in the  local search process. Therefore, the time for completing all the tasks does not increase with the iterations of the ITLTO algorithm. It is known that the completion time for all the tasks has a lower bound in such a network, and the objective function can not decrease infinitely. Therefore, the completion time for all the tasks will converge to a stable value after limited iterations.
\end{proof}\vspace{-1mm}

\textbf{Proposition 2:} The complexity of the proposed ITLTO algorithm is $O(N^2\times(\frac{S}{v_{max}^2}+1))$, where $S$ is the area of the task distribution range.
\begin{proof}
In UAV trajectory optimization, we limit the enumerated variable $T_n^f$ to the interval $[\frac{L}{v_{max}}, \frac{\epsilon L}{v_{max}}]$, where $L$ is the average distance between two sensing locations and $\epsilon$ is an adjustable parameter. Since we solve $T_n^f$ convex optimization problem for each $T_n^f$ enumerated, the complexity is $O(N\times\frac{L^2}{v_{max}^2})=O(N\times\frac{S}{v_{max}^2})$. In the UAV sensing location and sensing time optimization, the number of the local search processes is proportional to the number of tasks, i.e., the complexity is $O(N)$. The number of ITLTO iterations is relevant to the reduction of the completion time for all the tasks, which is directly proportional to the number of tasks $N$. Therefore, the complexity  is $O(N^2\times(\frac{S}{v_{max}^2}+1))$.
\end{proof}

\textbf{Proposition 3:} If the UAV flies from locations with lower sending rate to locations with higher sending rate, the optimal choice for the UAV is to increase sensing time to minimize the total completion time. Otherwise, the UAV should increase flying time to send more data in the flying process.
\begin{proof}
Without loss of generality, we take the trajectory from sensing locations $n-1$ to $n$ to analyze the relationship between UAV flying time and UAV sensing time. According to~(\ref{sensing time optimization}), we can have
\begin{equation}\vspace{-1mm}
    \delta_n \approx \frac{\theta R_s+Q(t_n)}{R(t_n)},
\end{equation}
where $\theta=C_n-1-\lfloor\frac{\ln\rho^{*}}{\ln\rho_n}\rfloor$ is constant if sensing location $n$ is fixed. Let $Q_0$ be the amount of data in the buffer when the UAV starts flying. If we increase $R_s$ by $\Delta R_s$, $Q_0$ will increase to $Q_0^{'}$ correspondingly, where $Q_0^{'} \approx Q_0(1+\frac{\Delta R_s}{R_s})$.

If the UAV flying time $T_{n-1}^f$ is fixed, the UAV needs to increase $\delta_n$ to guarantee the task completion probability of task $n$. The increment can be given by
\begin{equation}\vspace{-1mm}
\Delta \delta_n=\frac{\Delta R_s}{R(t_n)R_s}(\theta R_s+Q_0).
\end{equation}

If the UAV sensing time $\delta_n$ is fixed, the UAV needs to fly for more $\Delta T_{n-1}^f$ time slots to send data in order to guarantee the task completion probability. Let $Q(t_n)^{'}$ be the amount of data in the buffer when the UAV reaches sensing location $n$. The increased amount of data sent during the flying process can be expressed by $R_{ave}\Delta T_{n-1}^{f}=(Q_0^{'}-Q(t_n)^{'})-(Q_0-Q(t_n))$. According to~(\ref{task completion probability}), we have
\begin{equation}\vspace{-1mm}
\frac{Q(t_n)-R(t_n)\delta_n}{R_s}=\frac{Q(t_n)^{'}-R(t_n)\delta_n}{R_s+\Delta R_s},
\end{equation}
where $R_{ave}$ is the average sending rate on this segment of trajectory. Therefore, $\Delta T_{n-1}^f$ can be expressed by
\begin{equation}\vspace{-1mm}
\Delta T_{n-1}^f=\frac{\Delta R_s}{R_{ave}R_s}(\theta R_s+Q_0).
\end{equation}

It can be observed that if $R(t_n)\textgreater R_{ave}$, we have $\Delta \delta_n\textless \Delta T_{n-1}^f$, and thus the proposition is obtained.
\end{proof}

\vspace{-2mm}

\begin{figure*}[!t]
    \centering
    \subfigure[]{
    \includegraphics[width=4.5cm,height = 3.4cm]{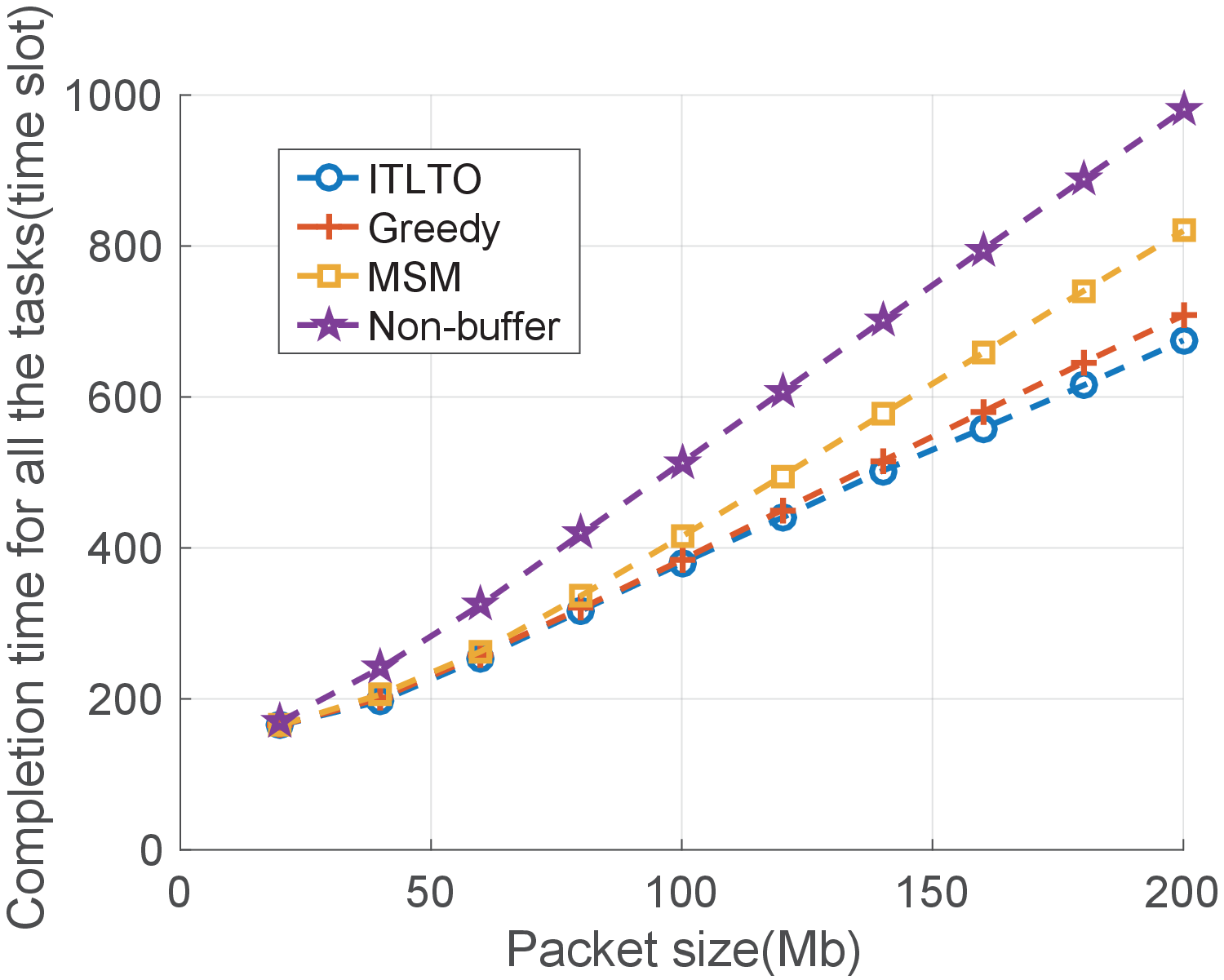}
    }\subfigure[]{
    \includegraphics[width=4.5cm,height = 3.4cm]{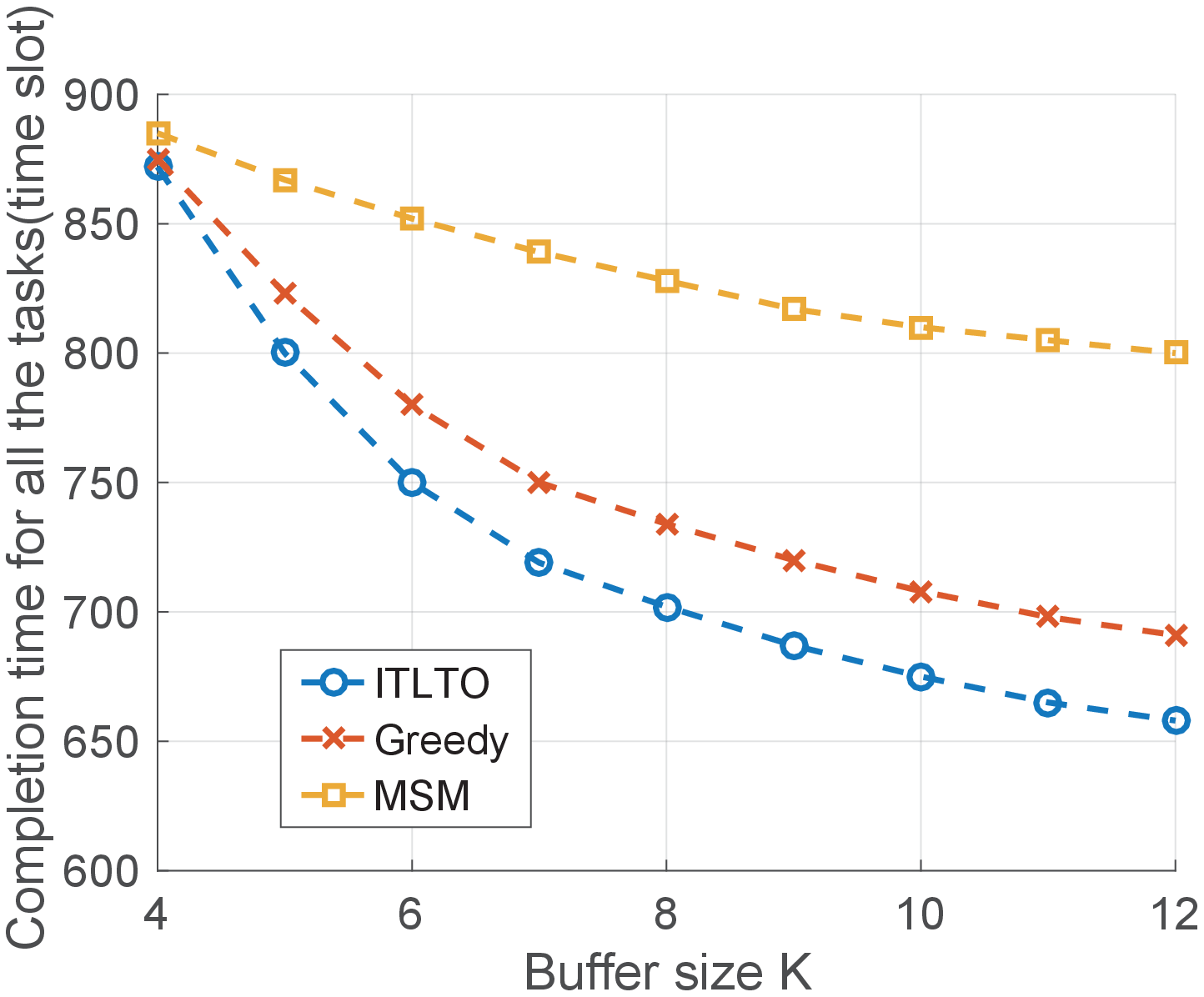}
    }
    \subfigure[]{
    \includegraphics[width=4.6cm,height = 3.2cm]{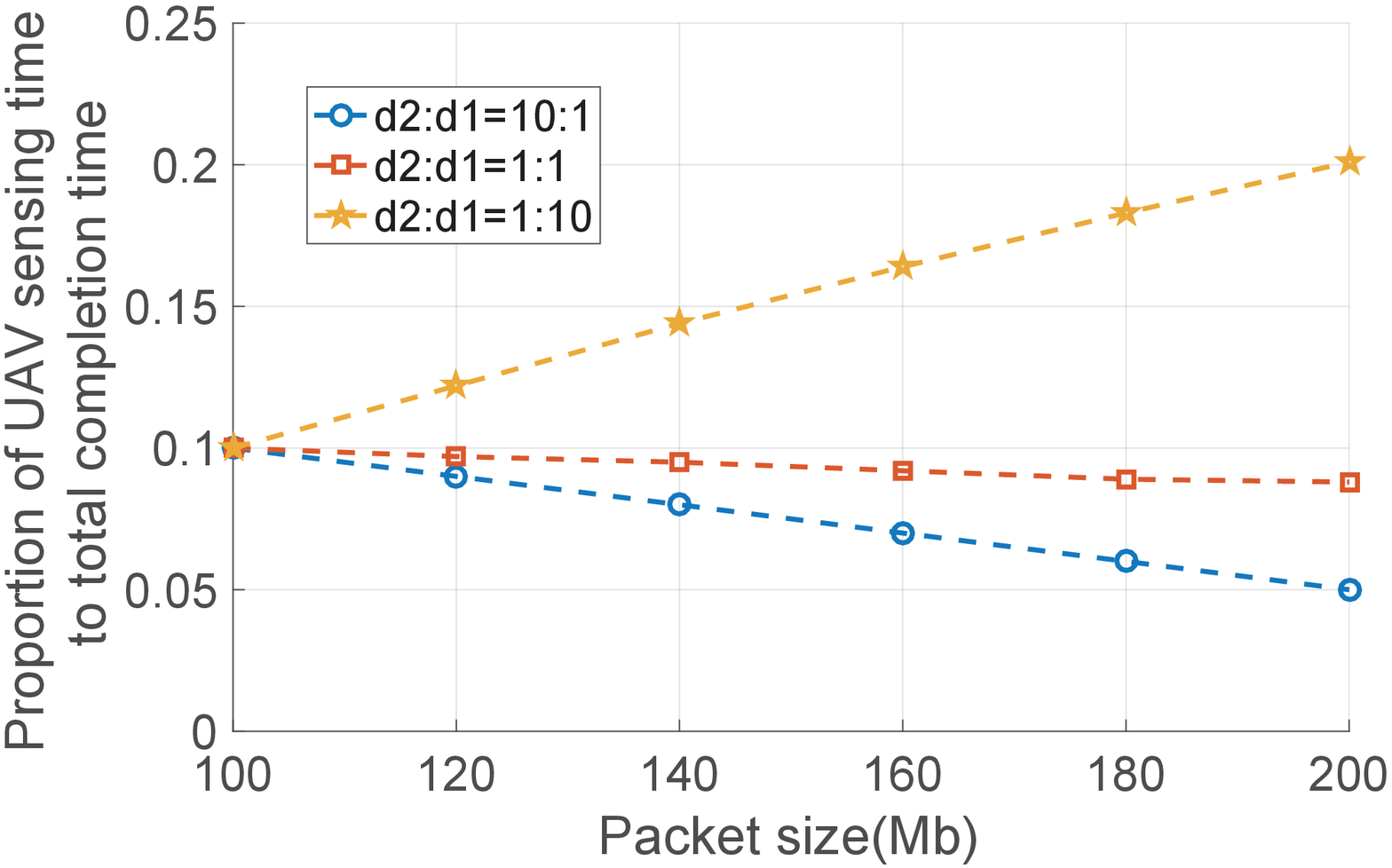}
    }
    \vspace{-3mm}
    \caption{ a) Completion time vs. packet size for different algorithms with $K = 10$. b) Completion time vs. buffer size for different algorithms with $R_s=200$Mb. c) Proportion of UAV sensing time to total completion time vs. packet size with $N=2$ and $K=10$.}
    \vspace{-7mm}
    \label{figure_2}
\end{figure*}

\vspace{-1mm}
\section{Simulation Results}
\vspace{-2mm}
\label{simulation}
In this section, we evaluate the performance of the proposed ITLTO algorithm. The selection of the simulation parameters is based on the existing specification~\cite{ZZHBS2018}. The number of tasks is set as $N=11$. We set the height of the BS as $H=20$ m, and the locations of tasks are randomly distributed on the ground of a 600m $\times$ 600m area. The UAV parameters are set as $P_T=23$ dBm and $v_{max}=20$ m/s. The transmission channel parameters are given as $W_B=1$ MHz, $\eta_{LoS}=1$, $\eta_{NLoS}=20$, $\alpha=12$, $\beta=0.135$, $L_{FS}=32.44$ and $\sigma^2=-96$ dBm. As for the parameters for UAV sensing, we set $\nu=0.1$, $C=1$, $\lambda=5$ and $p_{min}=0.9$.

For comparison, we also implement the following algorithms as benchmarks: 1) \textbf{Greedy method}: the UAV will select the location and speed which make the completion time of the current task minimized. 2) \textbf{Maximum speed method (MSM)}: the UAV performs sensing right over the tasks, and the UAV flies along line segments connecting sensing locations with speed being $v_{max}$. 3) \textbf{Non-buffer method}: The UAV needs to transmit all the sensory data before the next task. The UAV trajectory is optimized by the method introduced in~\cite{YJD2019}.

Fig. \ref{figure_2} (a) shows the relation between the completion time and the packet size $R_s$ with $K=10$. As $R_s$ exceeds $100$ Mb, the ITLTO method shows a better performance than MSM. The reason lies on that when the sensing rate is low, the data of one task stored in the buffer can be easily sent to the BS before the UAV reaches the next sensing location, and thus, the UAV tends to fly with $v_{max}$. However, when the sensing rate becomes higher, the burden on the buffer becomes heavier and the UAV has to decelerate and hover around locations with high sending rate. Besides, the ITLTO algorithm can obtain about $5\%$ gain compared to the greedy method when $R_s=200$Mb, since the greedy method does not consider the impact of the current task on the subsequent tasks. In the non-buffer method, the UAV needs to send all the data of a task before it executes the next task, which leads to an extra time cost.

Fig. \ref{figure_2} (b) shows the relation between the buffer size $K$ and the completion time $T$ using different algorithms with $R_s=200$Mb. It can be observed that the gap of the completion time between ITLTO and the two others increases with the buffer size.  It also shows that the completion time decreases with the buffer size. This is because a larger buffer helps improve the task completion probability, and thus reducing the UAV sensing time. This has justified that the usage of the buffer can decrease the completion time.

In Fig.~\ref{figure_2} (c), the proportion of UAV sensing time to the total completion time is plotted. In this simulation, we set $N=2$, and the distance between the task $i$ and the BS is denoted by $d_i$, respectively. Since the sending data rate is negatively related to distance, Fig.~\ref{figure_2} (c) also shows how the sending data rate influences the proportion of UAV sensing time. We can observe that when the UAV flies towards locations far from the BS, the ratio of UAV flying time increases with the packet size. When the UAV flies towards locations close to the BS, the proportion of UAV sensing time increases with the packet size. This is consistent with the analysis in Section \ref{analysis}.


\vspace{-2mm}
\section{Conclusions and Future Works}
\vspace{-2mm}
\label{conclusion}
In this letter, we have studied an Internet of UAVs aided with a limited buffer. We have proposed an iterative algorithm which contains UAV trajectory optimization, sensing location optimization, and sensing time optimization, to minimize the completion time for all the sensing tasks. From analysis and simulation results, we have two conclusions: 1) The completion time in a buffer-aided Internet of UAVs can be reduced by using a larger buffer. 2) When the packet size becomes larger, the proportion of the UAV sensing time  will increase if the sending rate is high, and otherwise the proportion of the UAV flying time grows. For future work, we will further investigate an online trajectory optimization algorithm, where fading and imperfect channel state information will be considered. Moreover, the security issue and the energy consumption can further be studied.
\vspace{-2mm}



\begin{thebibliography}{4}
\vspace{-2mm}
\bibitem{ZSH2019}
H. Zhang, L. Song, and Z. Han, \emph{Unmanned aerial vehicle applications over cellular networks for 5G and beyond}, Springer, 2019.


\bibitem{WQQ2018}
J. Gong, T. Chang, C. Shen, and X. Chen, ``Flight time minimization of UAV for data collection over wireless sensor networks," \emph{IEEE J. Sel. Areas Commun.}, vol. 36, no. 9, pp. 1942-1954, Sep. 2018.
\bibitem{YJD2019}
S. Zhang, H. Zhang, B. Di, and L. Song, ``Cellular cooperative unmanned aerial vehicles networks with sense-and-send protocol," \emph{IEEE Internet Things J.}, vol. 18, no. 2, pp. 1346-1359, Jan. 2019.

\bibitem{ZZZ2018}
Z. Zhou, J. Feng, B. Go, B. Ai, S. Mumtaz, J.Rodriguez, and M. Guizani, ``When mobile crowd sensing meets UAV: Energy-efficient task assignment and route planning," \emph{IEEE Trans. Commun.}, vol. 66, no. 11, pp. 5526-5538, Nov. 2018.

\bibitem{LH2017}
J. Li and Y. Han, ``Optimal resource allocation for packet delay minimization in multi-layer UAV networks," \emph{IEEE Commun. Lett.}, vol. 21, no. 3, pp. 580-583, Mar. 2017.

\bibitem{JWY2020}
X. Jiang, Z. Wu, Z. Yin, Z. Yang, and N. Zhao, ``Power consumption minimization of UAV relay in NOMA networks," \emph{IEEE Wireless Commun. Lett}, vol. 9, no. 5, pp. 666-670, May. 2020.

\bibitem{GH1985}
D. Gross and C. Harries, ``General arrival or service patterns," \emph{Fundamentals of queuing theory}, 5th ed. New Jersey, Wiley, 2018.
\bibitem{JM2011}
J. M. Smith, ``Properties and performance modeling of finite buffer M/G/1/K networks," \emph{Comput. Oper. Res.,} vol. 38, no. 4, pp. 740-754, Sep. 2011.
\bibitem{ZZD2019}
S. Zhang, H. Zhang, B. Di, and L. Song, ``Cellular UAV-to-X communications: Design and optimization for multi-UAV networks," \emph{IEEE Trans. Wireless Commun.,} vol. 18, no. 2, pp. 1346-1359, Jan. 2019.
\bibitem{ZZHBS2018}
S. Zhang, H. Zhang, Q. He, K. Bian, and L. Song, ``Joint power and trajectory optimization for UAV relay networks," \emph{IEEE Commun. Lett.,} vol. 22, no. 1, pp. 161-164, Jan. 2018.
\end{thebibliography}
\end{document}